\documentstyle[psfig,conf_iap,]{article}

\def\be{\begin{equation}}
\def\ee{\end{equation}}
\def\bea{\begin{eqnarray}}
\def\eea{\end{eqnarray}}

\begin{document}
\heading{%
%
Back Reaction of Cosmological Perturbations\\
and the Cosmological Constant Problem \\
%
} 
\par\medskip\noindent
\author{%
Robert H. Brandenberger$^{1}$
}
\address{%
Physics Department, Brown University, Providence, RI 02912, USA
}

\begin{abstract}
The presence of cosmological fluctuations influences the background
cosmology in which the perturbations evolve. This back-reaction
arises as a second order effect in the cosmological perturbation
expansion. The effect is cumulative in the sense that all
fluctuation modes contribute to the change in the background geometry,
and as a consequence the back-reaction effect can be large even if
the amplitude of the fluctuation spectrum is small. We review two 
approaches used to quantify back-reaction. In the first approach,
the effect of the fluctuations on the background is expressed in
terms of an effective energy-momentum tensor. We show that in the
context of an inflationary background cosmology, the long
wavelength contributions to the effective
energy-momentum tensor take the form of a negative cosmological
constant, whose absolute value increases as a function of time since
the phase space of infrared modes is increasing. This then leads to
the speculation that gravitational back-reaction may lead to a
dynamical cancellation mechanism for a bare cosmological constant,
and yield a scaling fixed point in the asymptotic future in
which the remnant cosmological constant satisfies $\Omega_{\Lambda} 
\sim 1$. We then discuss how infrared modes effect local observables
(as opposed to mathematical background quantities) and find that the
leading infrared back-reaction contributions cancel in single field
inflationary models. However, we expect non-trivial back-reaction of
infrared modes in models with more than one matter field.  
\end{abstract}
\section{Motivation}
It is well known that gravitational waves
propagating in some background space-time affect the dynamics of
the background. This back-reaction can be described in terms of an 
effective energy-momentum tensor $\tau_{\mu \nu}$. In the short wave limit, 
when the typical wavelength of the waves is small compared with the curvature 
of the background space-time, $\tau_{\mu \nu}$ has the form of a radiative 
fluid with an equation of state $p= \rho / 3$ (where $p$ and $\rho$ denote 
pressure and energy density, respectively).  

In most models of the early Universe, scalar-type metric perturbations are
more important than gravity waves. Here, we report on 
two studies of the back reaction problem for scalar gravitational 
perturbations. The first approach \cite{MAB96,ABM97} is
based on defining an effective energy-momentum tensor $\tau_{\mu \nu}$ which 
describes the back-reaction, and applying the results to an inflationary 
background cosmology. The second approach \cite{GG02} focuses on evaluating
the back-reaction of fluctuations on an observable measuring the local
Hubble expansion rate. Both studies are in the context of Einstein
gravity coupled to scalar field matter.

Our studies are closely related to work by Woodard and Tsamis 
\cite{WT1,WT2}
who considered the back-reaction of long wavelength gravitational waves
in pure gravity with a bare cosmological constant, and to work of Abramo
and Woodard \cite{Abramo1,Abramo2} who initiated the study of back-reaction of
infrared modes on local observables. 

In the following, we first review the derivation of the effective
energy-momentum tensor $\tau_{\mu \nu}$ which describes the back-reaction
of linear cosmological fluctuations on the background cosmology (Section 2),
and summarize the evaluation of this tensor in an inflationary cosmological
background (Section 3). We find that contribution of 
long wavelength (i.e. super-Hubble-scale) scalar metric fluctuations to 
$\tau_{\mu \nu}$ acts as a negative cosmological constant whose absolute
value increases in time since the phase space of infrared modes is increasing.
This gives rise to the speculations of Section 4 that back-reaction of
infrared modes may lead to a dynamical cancellation mechanism for the
bare cosmological constant. In Section 5 we address some major deficiencies
in this first approach to computing back-reaction. Many of these objections
were first raised by Unruh \cite{Unruh}. In Section 6 we then summarize
the present status of an improved approach to studying gravitational
back-reaction based on the calculation of corrections to an observable
measuring the local expansion rate. We find that the leading infrared
back-reaction effects vanish in single matter field models. However, based
on results concerning the parametric amplification of super-Hubble
cosmological fluctuations during inflationary reheating 
\cite{Bassett1,Fabio1,Bassett2,Fabio2}
we argue that in multi-field models the back-reaction of infrared modes
will be important.

\section{Framework}
The method of analyzing gravitational back-reaction \cite{MAB96} based on
the computation of an effective energy-momentum tensor for fluctuations 
is related to early work on the back-reaction of gravitational waves 
by Brill, Hartle and Isaacson \cite{Brill}, 
among others. The idea is to expand the Einstein equations to second order 
in the perturbations, to assume that the first order terms satisfy the 
equations of motion for linearized cosmological perturbations \cite{MFB92} 
(hence these terms cancel), to take the spatial average of the remaining 
terms, and to regard the resulting equations as equations for a new 
homogeneous metric $g_{\mu \nu}^{(0, br)}$ which includes the effect of 
the perturbations to quadratic order:
\be \label{breq}
G_{\mu \nu}(g_{\alpha \beta}^{(0, br)}) \, = \, 
8 \pi G \left[ T_{\mu \nu}^{(0)} + \tau_{\mu \nu} \right]\,
\ee
where the effective energy-momentum tensor $\tau_{\mu \nu}$ of 
gravitational back-reaction contains the terms resulting from spatial 
averaging of the second order metric and matter perturbations:
\be \label{efftmunu}
\tau_{\mu \nu} \, = \, < T_{\mu \nu}^{(2)} - 
{1 \over {8 \pi G}} G_{\mu \nu}^{(2)} > \, ,
\ee
where pointed brackets stand for spatial averaging, and the superscripts 
indicate the order in perturbation theory.

As analyzed in detail in \cite{MAB96,ABM97}, the back-reaction equation
(\ref{breq}) is covariant under linear space-time coordinate transformations
even though $\tau_{\mu \nu}$ is not invariant \footnote{See \cite{Unruh}, 
however, for important questions concerning the covariance of the analysis 
under higher order coordinate transformations.}. In the following, we will
work in longitudinal gauge (see e.g. \cite{MFB92} for a review of the
theory of cosmological perturbations). 

For simplicity, we shall take matter to be described in terms of a single 
scalar field. In this case, there is only one independent metric perturbation 
variable which we denote by $\phi(x,t)$, and in longitudinal gauge the 
perturbed metric can be written in 
the form
\be \label{metric}
ds^2 =  (1+ 2 \phi) dt^2 - a(t)^2(1 - 2\phi) \delta_{i j} dx^i dx^j  \, ,
\ee
where $a(t)$ is the cosmological scale factor. The energy-momentum tensor 
for a scalar field is 
\be 
T_{\mu \nu }=\varphi _{,\mu }\varphi _{,\nu }-g_{\mu \nu }\left[ {\frac 12}%
\varphi ^{,\alpha }\varphi _{,\alpha }-V(\varphi )\right] \,.
\ee

By expanding the Einstein tensor and the above energy-momentum tensor to 
second order in the metric and matter fluctuations $\phi$ and 
$\delta \varphi$, respectively, it can be shown that the non-vanishing 
components of the effective back-reaction energy-momentum tensor 
$\tau_{\mu \nu}$ become
\bea  \label{tzero}
\tau_{0 0} &=& \frac{1}{8 \pi G} \left[ + 12 H \langle \phi \dot{\phi} \rangle
- 3 \langle (\dot{\phi})^2 \rangle + 9 a^{-2} \langle (\nabla \phi)^2
\rangle \right]  \nonumber \\
&+& {1 \over 2} \langle ({\delta\dot{\varphi}})^2 \rangle + {1 \over 2} a^{-2} \langle
(\nabla\delta\varphi)^2 \rangle  \nonumber \\
&+& {1 \over 2} V''(\varphi_0) \langle \delta\varphi^2 \rangle + 2
V'(\varphi_0) \langle \phi \delta\varphi \rangle \quad ,
\eea
and 
\bea  \label{tij}
\tau_{i j} &=& a^2 \delta_{ij} \left\{ \frac{1}{8 \pi G} \left[ (24 H^2 + 16 
\dot{H}) \langle \phi^2 \rangle + 24 H \langle \dot{\phi}\phi \rangle
\right. \right.  \nonumber \\
&+& \left. \langle (\dot{\phi})^2 \rangle + 4 \langle \phi\ddot{\phi}\rangle
- \frac{4}{3} a^{-2}\langle (\nabla\phi)^2 \rangle \right] + 4 \dot{{%
\varphi_0}}^2 \langle \phi^2 \rangle  \nonumber \\
&+& {1 \over 2} \langle ({\delta\dot{\varphi}})^2 \rangle - {1 \over 6} a^{-2} \langle
(\nabla\delta\varphi)^2 \rangle - 
4 \dot{\varphi_0} \langle \delta \dot{\varphi}\phi \rangle  \nonumber \\
&-& \left. {1 \over 2} \, V''(\varphi_0) \langle \delta\varphi^2
\rangle + 2 V'( \varphi_0 ) \langle \phi \delta\varphi \rangle
\right\} \quad ,
\eea
where $H$ is the Hubble expansion rate.
\section{Application to Inflationary Cosmology}
The metric and matter fluctuation variables $\phi$ and $\delta \varphi$ are 
linked via the Einstein constraint equations, and hence all terms in the 
above formulas for the components of $\tau_{\mu \nu}$ can be expressed in 
terms of two point functions of $\phi$ and its derivatives. The two point 
functions, in turn, are obtained by integrating over all of the Fourier 
modes of $\phi$, e.g.
\be \label{tpf}
\langle \phi^2 \rangle \, \sim \,  \int_{k_i}^{k_u} {dk} k^2 \vert \phi_k
\vert^2 \, ,
\ee 
where $\phi_k$ denotes the amplitude of the k'th Fourier mode. The above
expression is divergent both in the infrared and in the ultraviolet. The
ultraviolet divergence is the usual divergence of a free quantum field
theory and can be ``cured" by introducing an ultraviolet cutoff $k_u$. In the
infrared, we will discard all modes $k < k_i$ with wavelength larger than 
the Hubble
radius at the beginning of inflation, since these modes are determined by
the pre-inflationary physics. We take these modes to contribute to the 
background.

At any time $t$ we can separate the integral in (\ref{tpf}) into the 
contribution of infrared and ultraviolet modes, the separation being 
defined by setting the physical wavelength equal to the Hubble radius. 
Thus, in an inflationary Universe the infrared phase space is continually 
increasing since comoving modes are stretched beyond the Hubble radius, 
while the ultraviolet
phase space is either constant (if the ultraviolet cutoff corresponds to a 
fixed physical wavelength), or decreasing (if the ultraviolet cutoff 
corresponds to fixed comoving wavelength). In either case, unless the 
spectrum of the initial fluctuations is extremely blue, two point 
functions such as (\ref{tpf}) will at later stages of an inflationary 
Universe be completely dominated by the infrared sector. In the following, 
we will therefore restrict our attention to this sector, i.e. to wavelengths 
larger than the Hubble radius.

In order to evaluate the two point functions which enter into the expressions 
for $\tau_{\mu \nu}$, we need to know the time evolution of the linear 
fluctuations $\phi_k$, which is given by the linear theory of cosmological 
perturbations \cite{MFB92}. On scales larger than the Hubble radius,
and for a time-independent equation of state, $\phi_k$ is constant in time.
The Einstein constraint equations relating the metric and matter fluctuations 
give 
\be 
\dot \phi +H\phi = 4\pi G\dot \varphi _0\,\delta \varphi \,.  \label{constr}
\ee
If the background scalar field $\varphi _0$ is rolling slowly, then  
$ \dot \varphi _0\simeq -{\frac{V^{\prime }}{3H}}$,
where a prime denotes the derivative with respect to the scalar matter field.
Thus,
\be 
\delta \varphi =-{\frac{2V}{V^{\prime }}}\,\phi \,.  \label{constr2}
\ee
Hence, in the expressions (\ref{tzero}) and (\ref{tij}) for $\tau_{\mu \nu}$, 
all terms with space and time derivatives can be neglected, and we obtain
\be 
\rho _{br}\equiv \tau _0^0\cong \left( 2\,{\frac{{V^{\prime \prime }V^2}}{{%
V^{\prime }{}^2}}}-4V\right) <\phi ^2>  \label{tzerolong}
\ee
and 
\be 
p_{br}\equiv -\frac 13\tau _i^i\cong -\rho_{br} \,,  \label{tijlong}
\ee

The main result which emerges from this analysis is that the equation of state
of the dominant infrared contribution to the energy-momentum tensor 
$\tau_{\mu \nu}$ which describes back-reaction takes the form of a 
{\it negative cosmological constant} 
\be \label{result}
p_{br}=-\rho _{br} \,\,\, {\rm with} \,\,\, \rho_{br} < 0 \, .
\ee

The second crucial result is that the magnitude of $\rho_{br}$ increases as
a function of time. This is due in part to the fact that, in an inflationary 
Universe, as time increases more and more wavelengths become longer than the 
Hubble radius and begin to contribute to $\rho_{br}$. 

How large is the magnitude of back-reaction? The basic point is that
since the amplitude of each fluctuation mode is small, we need a very
large phase space of infrared modes in order to induce any interesting
effects. In models with a very short period of primordial inflation,
the back-reaction of long-wavelength cosmological fluctuations hence 
will not be important. However, in many single field models
of inflation, in particular in those of chaotic inflation type \cite{Linde},   
inflation lasts so long that the infrared back-reaction effects can build up
to become important for the cosmological background dynamics. 

To give an example, consider chaotic inflation with a potential
\be 
V(\varphi )={\frac 12}m^2\varphi ^2\,.  \label{pot}
\ee
In this case, the values of $\phi_k$ for long wavelength modes are well 
known (see e.g. \cite{MFB92}), and the integral in (\ref{tpf}) can be
easily performed, thus yielding explicit expressions for the dominant terms
in the effective energy-momentum tensor. Comparing
the resulting back-reaction energy density $\rho_{br}$ 
with the background density $\rho_0$, we find
\be   \label{result2}
{\frac{{\rho_{br}(t)} }{{\rho_0}}} \, \simeq \, {\frac{{3} }{{4 \pi}}} {\frac{{
m^2 \varphi_0^2(t_i)} }{{M_P^4}}} \left[ {\frac{{\varphi_0 (t_i)} }{{\varphi_0
(t)}}} \right]^4 \, .
\ee
Without back-reaction, inflation would end \cite{Linde} when 
$\varphi_0 (t) \, \sim \, M_P$. Inserting this value into (\ref{result2}), 
we see that if 
\be 
\varphi_0 (t_i) \, > \, \varphi_{br} \, \sim \, m^{-1/3} M_P^{4/3} \, ,
\ee
then back-reaction will become important before the end of inflation and 
may shorten the period of inflation. It is interesting to compare this 
value with the scale 
$ \varphi_0 (t_i) \, \sim \, \varphi_{sr} \, = \, m^{-1/2} M_P^{3/2}$
above which the stochastic terms in the scalar field equation of motion
arising in the context of the stochastic approach to chaotic 
inflation \cite{Starob,Slava} are dominant. Notice that 
since $\varphi_{sr} \gg \varphi_{br}$
(recall that $m \ll M_P$), back-reaction effects can be very important in
the entire range of field values relevant to stochastic inflation.

\section{Speculations Concerning a Dynamical Relaxation Mechanism 
for $\Lambda$}

Since the back-reaction of cosmological fluctuations in an inflationary 
cosmology acts (see (\ref{result})) like a negative cosmological constant,
and since the magnitude of the back-reaction effect increases in time, one
may speculate \cite{RB98} that back-reaction will lead to a dynamical 
relaxation of the cosmological constant (see Tsamis \& Woodard \cite{WT1} 
for similar speculations based on
the back-reaction of long wavelength gravitational waves).

The background metric $g_{\mu \nu}^{(0, br)}$ including back-reaction evolves 
as if the cosmological constant at time $t$ were
\be \label{effcosm}
\Lambda_{\rm eff}(t) \, = \, \Lambda_0 + 8 \pi G \rho_{br}(t)
\ee
and not the bare cosmological constant $\Lambda_0$. Hence we propose to 
identify (\ref{effcosm}) with a time dependent effective cosmological 
constant. Since $\vert \rho_{br}(t) \vert$ increases as $t$ grows, the 
effective cosmological constant will decay. Note that even if the initial 
magnitude  of the perturbations is small, eventually (if inflation lasts a 
sufficiently long time) the back-reaction effect will become large enough to 
cancel any bare cosmological constant.

Furthermore, we speculate that this dynamical relaxation mechanism for 
$\Lambda$ will be self-regulating. As long as 
$\Lambda_{\rm eff}(t) \, > \, 8 \pi G \rho_m(t)$, where $\rho_m(t)$ stands 
for the energy density in ordinary matter and radiation, the evolution of 
$g_{\mu \nu}^{(0, br)}$ is dominated by $\Lambda_{\rm eff}(t)$. Hence, the 
Universe will be undergoing accelerated expansion, more scales will be 
leaving the Hubble radius and the magnitude of the back-reaction term will 
increase. However, once $\Lambda_{\rm eff}(t)$ falls below $\rho_m(t)$, the 
background will start to decelerate, scales will enter the Hubble radius, 
and the number of modes contributing to the back-reaction will decrease, 
thus reducing the strength of back-reaction. Hence, it is likely that there 
will be a scaling solution to the effective equation of motion for 
$\Lambda_{\rm eff}(t)$ of the form
\be \label{scaling}
\Lambda_{\rm eff}(t) \, \sim \, 8 \pi G \rho_m(t) \,.
\ee
Such a scaling solution would correspond to a contribution to the 
relative closure density of $\Omega_{\Lambda} \sim 1$.
\section{Criticism and Open Issues}
There are important concerns about the above formalism, and even more so
about the resulting speculations (some of these were first discussed in
print in \cite{Unruh}). On a formal level, since our back-reaction effect
is of second order in cosmological perturbation theory, it is necessary
to demonstrate covariance of the proposed back-reaction equation 
(\ref{breq}) beyond linear order, and this has not been done. Next, it might
be argued that by causality super-Hubble fluctuations cannot affect local
observables. Thirdly, from an observational perspective one is not
interested in the effect of fluctuations on the background metric (since
what the background is cannot be determined precisely using local 
observations). Instead, one should compute the back-reaction of cosmological
fluctuations on observables describing the local Hubble expansion rate. One
might then argue that even if long-wavelength fluctuations have an effect
on the background metric, they do not influence local observables.
Finally, it is clear that the speculations in the previous section involve
the extrapolation of perturbative physics deep into the non-perturbative
regime. 

These important issues have now begun to be addressed. Good physical
arguments can be given \cite{Abramo1,Abramo2} supporting the idea that
long-wavelength fluctuations can effect local physics. Consider, for
example, a black hole of mass $M$ absorbing a particle of mass $m$. Even
after this particle has disappeared beyond the horizon, its gravitational
effects (in terms of the increased mass of the black hole) remain
measurable to an external observer. A similar argument can be given in
inflationary cosmology: consider an initial localized mass fluctuation
with a characteristic physical length scale $\lambda$
in an exponentially expanding background. Even after the length scale
of the fluctuation redshifts to be larger than the Hubble radius, the
gravitational potential associated with this fluctuation remains measurable.
On a more technical level, it has recently been shown that super-Hubble
scale (but sub-horizon-scale) metric fluctuations can be parametrically
amplified during inflationary reheating 
\cite{Bassett1,Fabio1,Bassett2,Fabio2}. 
This clearly demonstrates a coupling between local physics and
super-Hubble-scale fluctuations.

These arguments, however, make it even more important to focus on
back-reaction effects of cosmological fluctuations on local physical
observables rather than on the mathematical background metric. This topic
will be discussed in the following section. The results of that analysis
will then determine the answer to the first of the concerns listed at the
beginning of this section.

It is obvious that even in models in which the perturbative back-reaction
results of Sections 2 - 3 have locally measurable implications, the
analysis has to be extended beyond perturbation theory to justify the
speculations of Section 4. For some initial ideas in this direction, see
\cite{WT3,AWT}.  

\section{Back-Reaction on Local Observables}
In this section we will be summarizing recent work \cite{GG02} in which
the leading infrared back-reaction effects on a local observable
measuring the Hubble expansion rate were calculated. Note that initial
work on gravitational back-reaction effects of infrared modes on local
observables was done in \cite{Abramo1,Abramo2}, using different methods
and a different observable than the one used below.

If we consider a perfect fluid with velocity four vector $u^{\alpha}$
in an inhomogeneous cosmological geometry, the local expansion rate which
generalizes the Hubble expansion rate $H(t)$ of homogeneous isotropic
Friedmann-Robertson-Walker cosmology is given by ${1 \over 3} \Theta$,
where $\Theta$ is the four divergence of $u^{\alpha}$:
\be
\Theta \, = \, u^{\alpha}_{; \alpha} \, ,
\ee
the semicolon indicating the covariant derivative.
In \cite{GG02}, the effects of cosmological fluctuations on this variable
were computed to second order in perturbation theory. To leading order
in the infrared expansion, the result is
\be \label{brresult1}
\Theta \, = \, 3 {{a^{\prime}} \over {a^2}}
\bigl( 1 - \phi + {3 \over 2} \phi^2 \bigr) - 3 {{\phi^{\prime}} \over a} \, ,
\ee
where the prime denotes the derivative with respect to conformal time.
If we now calculate the spatial average of $\Theta$, the term linear in
$\phi$ vanishes, and - as expected - we are left with a quadratic back-reaction
contribution.

Superficially, it appears from (\ref{brresult1}) that there is a 
non-vanishing back-reaction effect at quadratic order which is not
suppressed for super-Hubble modes. However, we must be careful and
evaluate $\Theta$ not at a constant value of the background coordinates,
but rather at a fixed value of some physical observable. For example,
if we work out the value of $\Theta$ in the case of a matter-dominated
Universe, and express the result as a function of the proper time
$\tau$ given by
\be
d \tau^2 \, = \, a(\eta)^2 (1 + 2 \phi) d \eta^2
\ee
instead of as a function of conformal time $\eta$, then we find that the
leading infrared terms proportional to $\phi^2$ exactly cancel, and that
thus there is no un-suppressed infrared back-reaction on the local
measure of the Hubble expansion rate.

A more relevant example with respect to the discussion in earlier sections
is a model in which matter is given by a single scalar field. In this
case, the leading infrared back-reaction terms in $\Theta$ are again given by
(\ref{brresult1})
which looks different from the background value $3 H$. However, once again
it is important to express $\Theta$ in terms of a physical background variable.
If we choose the value of the matter field $\varphi$ as this variable,
we find after easy manipulations that, including only the leading
infrared back-reaction terms, 
\be
\Theta ( \varphi ) \, = \, \sqrt{3} \sqrt{V(\varphi)} \, . 
\ee
Hence, once again the leading infrared back-reaction contributions vanish,
as already found in the work of \cite{Abramo2} which considered the
leading infrared back-reaction effects on a local observable different than
the one we have used, and applied very different methods \footnote{For a
different approach which also leads to the conclusion that there can be
no back-reaction effects from infrared modes on local observables in
models with a single matter component see \cite{Afshordi}.}.

However, in a model with two matter fields, it is clear that if we
e.g. use the second matter field as a physical clock, then the leading
infrared back-reaction terms will not cancel in $\Theta$, and that thus
in such models infrared back-reaction will be physically observable. The
situation will be very much analogous to what happens in the case of
parametric resonance of gravitational fluctuations during inflationary
reheating. This process is a gauge artifact in single field models of
inflation \cite{Fabio1} (see also \cite{Parry,Zhang}), 
but it is real and unsuppressed in certain
two field models \cite{Bassett2,Fabio2}. In the case of two field
models, work on the analysis of
the back-reaction effects of infrared modes on the observable representing
the local Hubble expansion rate is in progress. 
\section{Conclusions}
We have summarized the present status of the work on the gravitational
back-reaction of cosmological fluctuations. In an inflationary background
cosmology, the perturbations generated during inflation are shown to
contribute as an effective energy-momentum tensor of the form of a negative
cosmological constant to the evolution of the background metric. In
addition, the absolute value of this induced cosmological constant grows
in time since the phase space of infrared modes is increasing. This leads
to the intriguing speculation that gravitational back-reaction of
fluctuations may provide a dynamical cancellation mechanism for the
cosmological constant, leaving behind a remnant effective cosmological
constant which at all sufficiently late times corresponds to 
$\Omega_{\Lambda} \sim 1$.

However, we have also shown that in single field models, the leading infrared
terms in the back-reaction equation cancel when calculating the local 
Hubble expansion rate as a function of physical variables. We
have argued why in two-field models we do not expect this cancellation to
persist.

Obviously, a lot more work is required in order to be able to extend the
present calculations, which show at - leading order in the cosmological 
perturbation expansion - the onset of dynamical relaxation of $\Lambda$, to
higher orders and to a full nonlinear argument.

\acknowledgements{
The results summarized here are based on work done in collaboration with
R. Abramo, G. Geshnizjani and V. Mukhanov. I wish to thank them for
sharing their insights. I also wish to thank R. Abramo, A. Guth, A. Linde,
W. Unruh, N. Tsamis and in particular R. Woodard for important discussions
which heavily influenced the ideas presented in the last two sections of
this contribution. I also wish to thank the organizers of this Colloque
for inviting me to present this work, and for their wonderful hospitality
at the IAP. This work was supported in part by the US Department of Energy
under Contract DE-FG0291ER40688, Task A.}

\vfill
\end{document}